**OPEN ACCESS**

# Dashboard Task Monitor for Managing ATLAS User Analysis on the Grid

To cite this article: L Sargsyan *et al* 2014 *J. Phys.: Conf. Ser.* **513** 032083

View the article online for updates and enhancements.

## Related content

- ATLAS job monitoring in the Dashboard Framework
  J Andreeva, S Campana, E Karavakis et al.

- Experiment Dashboard - a generic, scalable solution for monitoring of the LHC computing activities, distributed sites and services
  J Andreeva, M Cinquilli, D Dieguez et al.

- Common Accounting System for Monitoring the ATLAS Distributed Computing Resources
  E Karavakis, J Andreeva, S Campana et al.

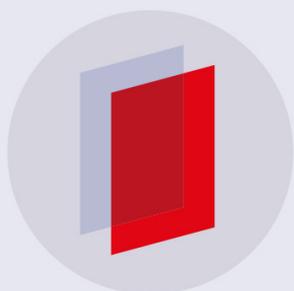

# IOP ebooks™

**Bringing you innovative digital publishing with leading voices to create your essential collection of books in STEM research.**

Start exploring the collection - download the first chapter of every title for free.

# Dashboard Task Monitor for Managing ATLAS User Analysis on the Grid

**L Sargsyan[1], J Andreeva[2], M Jha[3], E Karavakis[2], L Kokoszkiewicz[4], P Saiz[2], J Schovancova[5], D Tuckett[2] on behalf of the ATLAS Collaboration**

[1] A I Alikhanyan National Scientific Laboratory, Yerevan, Republic of Armenia
[2] CERN, European Organization for Nuclear Research, Switzerland
[3] Purdue University, United States of America
[4] w3widgets.com, Poland
[5] Brokhaven National Laboratory, Upton, United States of America

E-mail: Laura.Sargsyan@cern.ch

**Abstract**. The organization of the distributed user analysis on the Worldwide LHC Computing Grid (WLCG) infrastructure is one of the most challenging tasks among the computing activities at the Large Hadron Collider. The Experiment Dashboard offers a solution that not only monitors but also manages (kill, resubmit) user tasks and jobs via a web interface. The ATLAS Dashboard Task Monitor provides analysis users with a tool that is independent of the operating system and Grid environment. This contribution describes the functionality of the application and its implementation details, in particular authentication, authorization and audit of the management operations.

## 1. Introduction

The Worldwide LHC Computing Grid (WLCG) [1] infrastructure is set up to process the data from the experiments at the Large Hadron Collider located at CERN. ATLAS [2], one of the biggest LHC experiments, produces a huge amount of data. Thousands of scientists analyze this data in search of new particles in the head-on proton-lead collisions. More than 350000 ATLAS analysis jobs are submitted daily on the Grid. This number is steadily growing [3]. Reliable and flexible monitoring applications are required to follow the job processing. In such an environment users need to be able to monitor their jobs in real-time and to kill or to resubmit them if something goes wrong. The Experimental Dashboard [4] monitoring framework that was developed for the LHC experiments provides a solution for ATLAS analysis users - the Dashboard Analysis Task Monitor.

The Experimental Dashboard Task Monitor application discussed in this article is a web-based tool that enables ATLAS users to track progress of the job processing in detail and to manage them.
The main focus of this paper is a description of the security model of the application and its implementation details.

Access to the application is granted only to users with a valid Grid certificate. All parameters passed to the server side are sanitized and protected against cross-site scripting (XSS) [5] and cross-site request forgery (CSRF) [6] attacks. The audit information about all kill requests is stored in the local log file, CERN Central Security Logging server and Dashboard Central Repository.

## 2. Task Monitoring Architecture

ATLAS physicists use the PanDA [7] workload management system for job processing. A small percentage of ATLAS user jobs are still submitted through GANGA [8] to the gLite WMS [9]. Dashboard Analysis Task Monitoring application collects and exposes information that describes the progress of the user task processing. It uses Dashboard Data Repository (ORACLE) as a backend. Dashboard collectors consume job monitoring information from the PanDA job processing database, from jobs submitted through GANGA to WMS or local batch systems, while monitoring information is collected from the ActibeMQ Message brokers [10].

The main components of the Experimental Dashboard framework are information collectors, data repository, and services that retrieve and expose monitoring data. The comprehensive description of the Experimental Dashboard framework components is provided in [11].
ATLAS job monitoring architecture is presented in Figure 1.

All the collected information is exposed to the user via the web User Interface (UI).

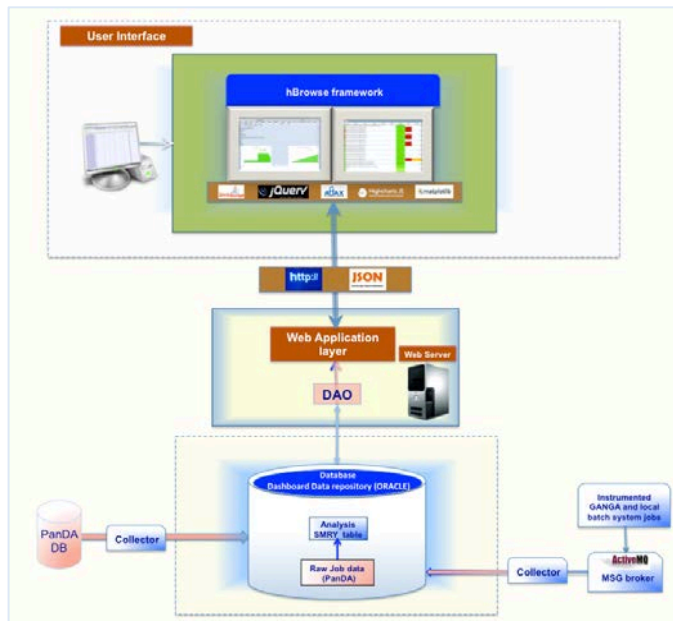

Figure 1. Architecture of the ATLAS Task Monitor application.

## 3. User Interface

### 3.1. Implementation

The user web interface is implemented using the hBrowse [12] visualization framework. hBrowse is a client side framework, which communicates with the server using AJAX requests. Server and client side separation allows the client or server side implementation to be changed independently of each other.

hBrowse uses client-side model–view–controller (MVC) architectural pattern (Figure 2). The model is a JavaScript object. The controller initializes the model and the view, synchronizes the URL #hash with the model and updates the view. The model holds the view state and application data (cached data). The view manages UI controls, tables and plots and pushes UI control changes to the model.

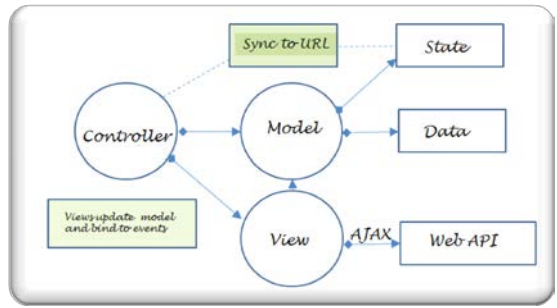

Figure 2. Client-side MVC components.

The hBrowse framework uses jQuery [13] and many of its plug-ins such as BBQ (Back Button and Query Library), Highcharts, DataTables, LiveSearch, etc.

3.2. Functionality
The User Interface provides access to the user's tasks (collection of jobs based on output container) and jobs using a secure web connection (HTTPS) and Grid Certificate. There are two visualization modes: View mode and Manage mode. In the View mode a user can view his/her jobs and also jobs of other ATLAS colleagues. The Manage mode provides the ability for the task owner to kill:
- all jobs in a task
- all jobs in a task running on a given site
- a specific job or set of jobs

The User Interface can display a list of tasks submitted over a chosen time range ( lastDay, last2Days, etc. ) or for a specified time period (From .. To). The task meta information such as the last modification time, the input dataset, sites where jobs of this particular task are running, etc. could be accessed by clicking on the "+" symbol next to the "Graphically" column. The snapshot of the Dashboard Task Monitor UI is presented in Figure 3.

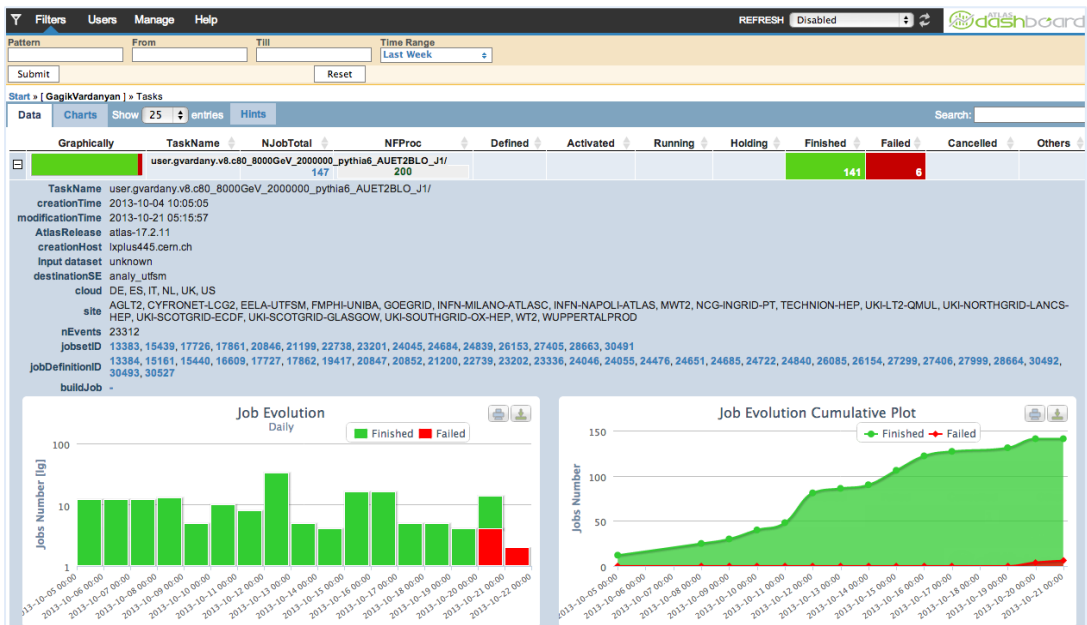

Figure 3. The User Interface.

Users can check the status of the jobs belonging to the chosen task and investigate the reason of failures, the resubmission history, etc. Jobs could be filtered by job status and site(s), where jobs of the given task are running. A wide variety of graphical plots is available at the task and job level. They help users to manage tasks. E.g. user can identify a problematic site using 'Jobs distributed by site' plot, kill job(s) directly from the UI and then resubmit jobs to another site.

The User Interface provides also on-the-fly filtering, sorting, sorting column(s) highlighting, variable length pagination as well as full bookmarking capability, working "refresh" capability, "breadcrumbs" navigation, etc.

## 4. Security model implementation

### 4.1. Authentication
Authentication permissions are required for viewing monitoring data (View mode) and for managing user tasks (killing) in Manage mode. In View mode a client could view his/her tasks/jobs and also tasks/jobs of other ATLAS analysis users. Managing of user jobs is allowed only by the job owner.

Access to the application is performed using a secure web connection (HTTPS) and Grid certificate. Authentication by X509 Grid certificate is mandatory and performed entirely within the front-end server. The certificate-based authentication requires SSL client verification. Optional client verification is enabled at a global server level for all HTTPS connections so the method uses these results.

### 4.2. Manage mode
The Manage mode actions are presented in Figure 4 and described below.

*4.2.1. Handling of the session id.* If authentication succeeds and Manage mode is chosen, processing is allowed to continue with the next step. On this step a session id is generated. Session id information is embedded within the form as a hidden field during the client request and submitted with the HTTP POST command. An important aspect of managing state within the web application is the "strength" of the session id itself. The generation of the session id should fulfil criteria: it should be random, unpredictable, and cannot be reproduced. In order to meet these requirements, the application utilises a strong method to generate a session id.

The special collector inserts the session id in the dashboard central repository. The session information is time limited. It expires after a specified timeout period. The ORACLE procedure revokes the session id when a threshold has been reached.

*4.2.2. Implementation of killing functionality.* The "kill" procedure sanitizes all parameters which are passed during request to prevent embedding of malicious JavaScript, VBScript, ActiveX, HTML, or Flash by an attacker [4]. Performing the appropriate validation provides protection against malicious client-side scripts and any cross-site scripting forgery attacks.

To avoid SQL injection [14] vulnerabilities the application uses prepared statements and bind variables.

*4.2.2.1. Authorization.* Only the owner of the job should be allowed to kill or to resubmit his/her jobs. During this step the procedure gathers information associated with the authenticated user and checks the local policy (one can kill only his/her jobs). If the distinguished name (DN) of the requestor and task/job owner is identical the access is granted and the "killJobs" request is sent to the PanDA server. Otherwise processing terminates with the error message.

*4.2.2.2. Audit.* The audit logging data is stored in a log file locally on the server, on the CERN Central Security logging, and in the dashboard central repository. Each message contains the following information: client IP address, passed parameters, client DN, and PanDA server replay message. This data is used by the users support team to review the results, identify and fix the problem.

Monitoring and review of this data, as determined by the criticality of the application, past experience with incidents, and general risk assessment is important.

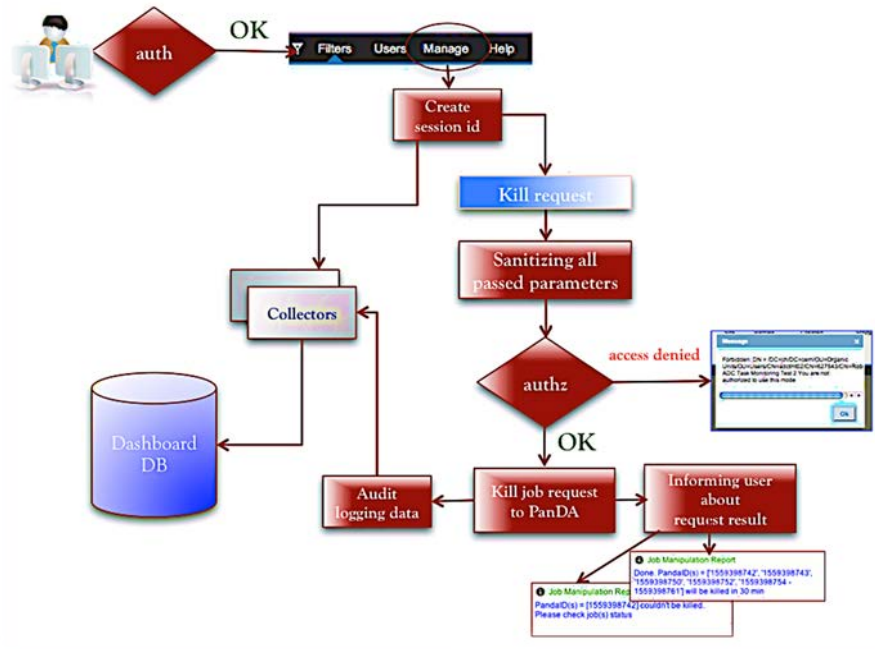

Figure 4. Manage mode actions.

## 5. Conclusions

The Dashboard Task monitor provides analysis users with the ability to manage and view their tasks using web browser regardless of the operating system and Grid environment. It offers a complete and detailed view of user tasks, detailed job information with full resubmission history. The application has become more interactive, as it supports cancellation. The next steps consist of enabling ability to resubmit failed jobs from the UI.

The kill job functionality was tested by pilot users and CERN security experts. It proved to be reliable from a security point of view.

An attractive, intuitive web interface with a wide selection of graphical plots, and the ability to manage jobs from the web UI makes the application more popular among experienced and new ATLAS analysis users.

## 6. Acknowledgment

The authors are thankful to Dario Barberis, Douglas Benjamin, Simone Campana, Andres Pacheco Pages for many helpful suggestions and support. We are particularly grateful for web application security guidance and vulnerability checks performed by Sebastian Lopienski.